\documentclass[a4paper,10pt,twoside]{cpc-hepnp}
\usepackage{CJK,upgreek,fancyhdr}
\usepackage{multicol}
\usepackage{graphicx}
\usepackage{booktabs}
\usepackage{amssymb,bm,mathrsfs,bbm,amscd}
\usepackage[tbtags]{amsmath}
\usepackage{lastpage}
\usepackage{color}

\begin{document}
\begin{CJK*}{GBK}{song}

\fancyhead[c]{\small Chinese Physics C~~~Vol. xx, No. x (201x) xxxxxx}
\fancyfoot[C]{\small 010201-\thepage}

\footnotetext[0]{Received 31 June 2015}

\title{PMT overshoot study for the JUNO prototype detector\thanks{Supported by Strategic Priority Research Program A-JUNO,High Energy Physics Experiment and Detector R\&D and National Natural Science Foundation of China  }}

\author{%
      Feng-Jiao Luo$^{1,2,3;1)}$\email{luofj@ihep.ac.cn}%
\quad Yue-Kun Heng$^{1,2;2)}$\email{hengyk@ihep.ac.cn}%
\quad Zhi-Min Wang$^{2;3)}$\\\email{wangzhm@ihep.ac.cn}%
\quad Pei-Liang Wang$^{1,2}$
\quad Zhong-Hua Qin$^{1,2}$
\quad Mei-Hang Xu$^{1,2}$\\
\quad Dong-Hao Liao$^{2,4}$
\quad Hai-Qiong Zhang$^{1,2,3}$
\quad Yong-Bo Huang$^{2,3}$\\
\quad Xiang-Cui Lei$^{1,2,3}$
\quad Sen Qian$^{1,2}$
\quad Shu-Lin Liu$^{1,2}$\\
\quad Yuan-Bo Chen$^{1,2}$
\quad Yi-Fang Wang$^{1,2}$
}
\maketitle

\address{%
$^1$ State Key Laboratory of Particle Detection and Electronics, Beijing 100049, China\\
$^2$ Institute of High Energy Physics, Chinese Academy of Sciences, Beijing 100049, China\\
$^3$ University of Chinese Academy of Science, Beijing, China\\
$^4$ Guangxi University, Nanning 530004, China\\
}

\begin{abstract}
The quality of PMT signals is a key for large-size and high-precision neutrino experiments, while most of these experiments are affected by the overshoot of PMT signal from the positive HV-single cable scheme. Overshoot affects the trigger, dead time and charge measurement from a detector. For the JUNO prototype detector, we have performed a detailed study and calculation on PMT signal overshoot to control the ratio of overshoot to signal amplitude to $\sim$1\%, with no effect on other PMT parameters.
\end{abstract}



\footnotetext[0]{\hspace*{-3mm}\raisebox{0.3ex}{$\scriptstyle\copyright$}2013
Chinese Physical Society and the Institute of High Energy Physics
of the Chinese Academy of Sciences and the Institute
of Modern Physics of the Chinese Academy of Sciences and IOP Publishing Ltd}%

\begin{multicols}{2}

\section{Introduction}

The JUNO\cite{lab1} project is proposed to determine the neutrino mass hierarchy using a 20 kton underground liquid scintillator detector. As a multipurpose underground neutrino observatory, JUNO will also measure neutrino oscillation parameters to better than 1\% accuracy and measure neutrinos or antineutrinos from terrestrial and extra-terrestrial sources. It will provide exciting opportunities to address important topics in neutrino and astro-particle physics, including supernova bursts, the diffuse supernova neutrino background, geoneutrinos, atmospheric neutrinos and solar neutrinos. According to its preliminary design, JUNO will distribute $\sim$17000 $20^{''}$ PMTs to reach better than 3\% energy resolution at 1~MeV\cite{lab2}.

Following the JUNO schedule and R\&D requirements, especially for the newly developed high quantum efficiency and large area PMTs\cite{lab3}, a prototype detector of JUNO is planned at the Institute of High Energy Physics (IHEP) Chinese Academy of Science, China\cite{lab4}. The prototype detector is designed with 51 PMTs from three companies: $8^{''}$ dynode PMTs and $20^{''}$ dynode PMTs from Japan HAMAMATSU, $8^{''}$ and $20^{''}$ Micro-Channel Plate ( MCP ) PMT from Nanjing Night Vision Technology Co. ( NNVT ) and $9^{''}$ dynode PMT from Hainan Zhanchuang Photonics Co. ( HZC ). The positive HV scheme is used on these PMTs, and the 50-Ohm coaxial cable is carrying both the HV and signals. The scheme is widely used in many other experiments, such as Daya Bay\cite{lab5}, Borexino\cite{lab6}, Chooz\cite{lab7}, and Double Chooz\cite{lab8} experiments.

A simplified schematic of the positive HV-single cable scheme is shown in Fig.\ref{fig1}, where a capacitor is used as decoupler to separate the signal and high voltage. This scheme has many advantages such as lower PMT noise, fewer cable connection and lower cost, however the overshoot following a signal also causes problems for charge measurements and system triggering as shown by the Double Chooz\cite{lab9}, KamLAND\cite{lab10}, SNO\cite{lab11}, Borexino\cite{lab12}, and Daya Bay experiments: it makes the trigger system less efficient at detecting lower energy events and distorts the charge measurement of signals adjacent in time or after large signals from a muon crossing the detector. In order to overcome the disadvantages of overshoot, much work has been done by different experiments: two data dead time monitor systems were installed to determine the inefficiency from overshoot in Double Chooz\cite{lab13}; specific triggering or data acquisition systems were developed in SNO, KamLAND and Borexino. At the same time, we also do not find a common strategy for handling overshoot among these experiments. In this paper, we show a detailed study of the overshoot ratio and how to control overshoot through the optimization within the PMT HV divider and HV-signal decoupler.
\begin{center}
\includegraphics[width=8cm]{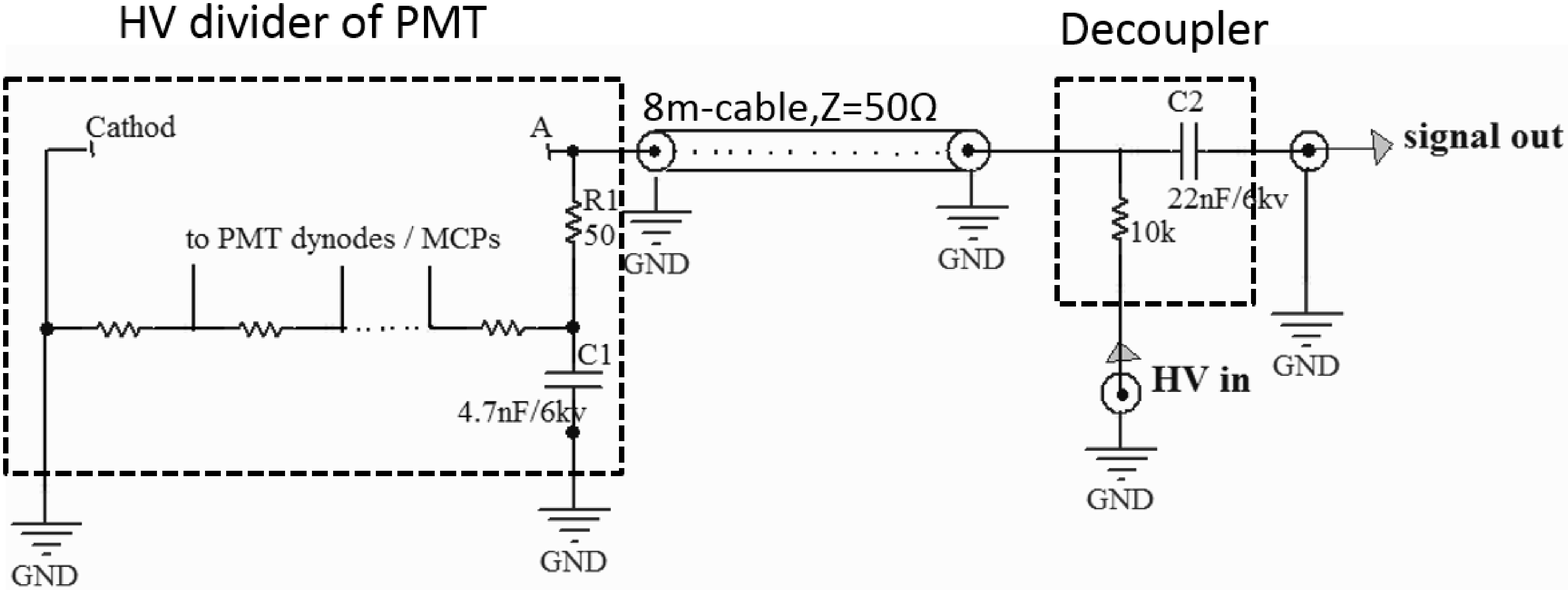}
\figcaption{Simplified schematic diagram of PMT, decoupler, output signal and positive high voltage}
\label{fig1}
\end{center}

\section{The overshoot}
A schematic of the PMT test system with positive HV and single cable is shown in Fig.\ref{fig2}. A LED was used to fire PMT and PMT waveforms were sampled by oscilloscope.
\begin{center}
\includegraphics[width=7cm]{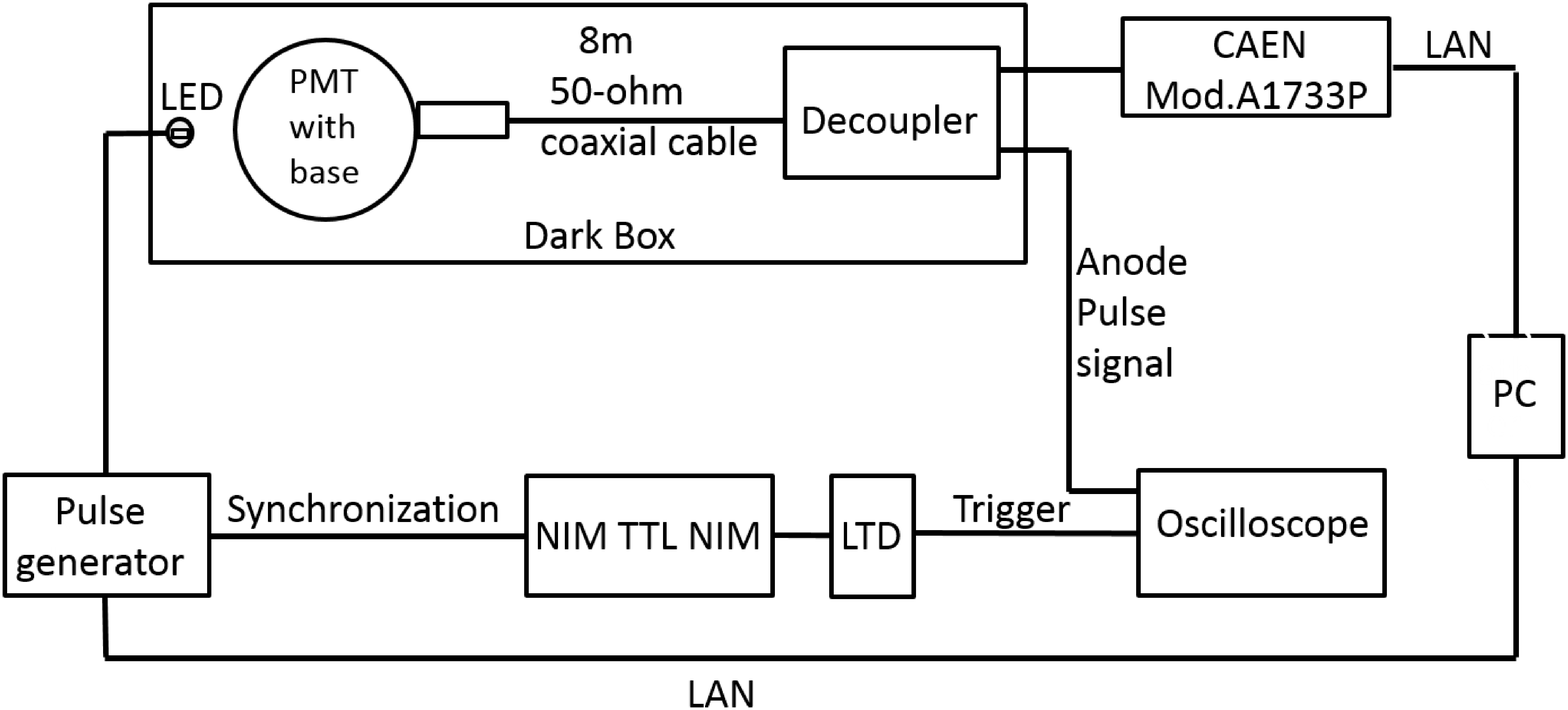}
\figcaption{The schematic view of PMT testing system with positive HV and one cable }
\label{fig2}
\end{center}

The PMT and LED were put in a dark box, where the LED was driven by a program controlled pulse-generator. At the same time, a synchronized gate from the pulse generator was sent to a low threshold discrimination ( LTD ) to trigger the oscilloscope. The high voltage was provided by a CAEN Mod.A1733P in CAEN SY4527 crate, and the PMT anode signal after decoupler is sent to an oscilloscope with sampling rate of 1~GHz. Some sampled PMT waveforms are shown in Fig.\ref{fig3}, where the recovery time is defined from the rising edge of PMT pulse to the overshoot amplitude recovered to 0.5\% of maximum amplitude of signal and the ratio of overshoot to signal amplitude is $\sim$10.8\%, which would be a major problem for charge measurements of signals adjacent in time as shown in Daya Bay\cite{lab14}. It is important to minimize the overshoot as much as possible.
\begin{center}
\includegraphics[width=8cm]{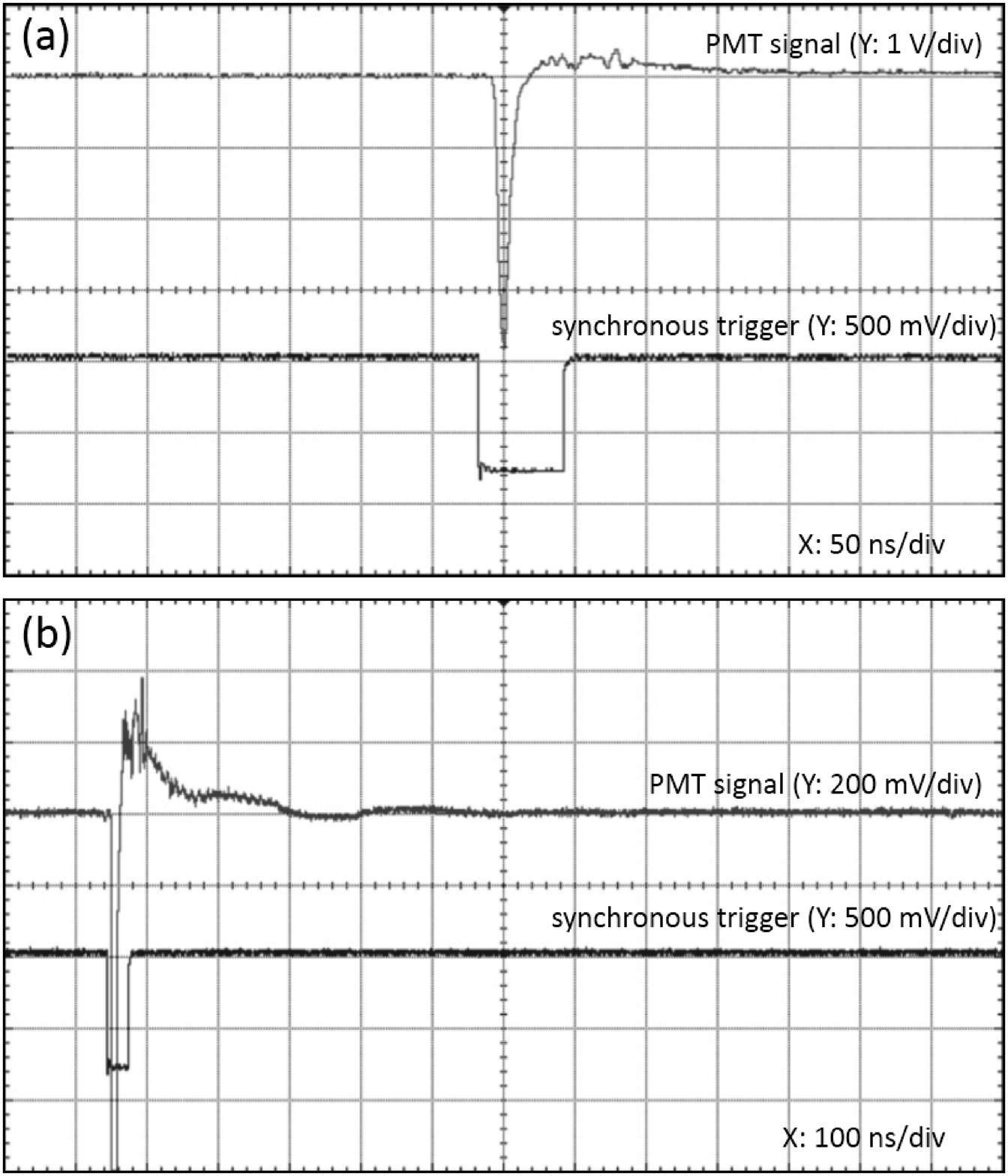}
\figcaption{PMT signal after decoupler: (a)1~V/div and 50~ns/div, (b) the same signal under different scales, 200~mV/div and 100~ns/div. The signal amplitude is about 3.7~V. Overshoot is clear and the amplitude of overshoot is about 400~mV or 10.8\%, it takes 700~ns to recover.}
\label{fig3}
\end{center}

\section{Circuit optimization}

We further simplified the positive HV PMT + decoupler + oscilloscope circuit as shown in Fig.\ref{fig4}, where C2 is the capacitor in the decoupler and R2 is the electrical load. The charge collected by the PMT anode will be released through the capacitors of C1 in the PMT HV divider and C2, while the discharge of C1 and C2 will affect overshoot. In the following sections, we will measure the relationship among overshoot and $C1\times R1$ or $C2\times R2$, model the overshoot, and optimize the circuit to minimize overshoot to meet the requirement of the JUNO prototype.
\begin{center}
\label{fig4}
\includegraphics[width=5cm]{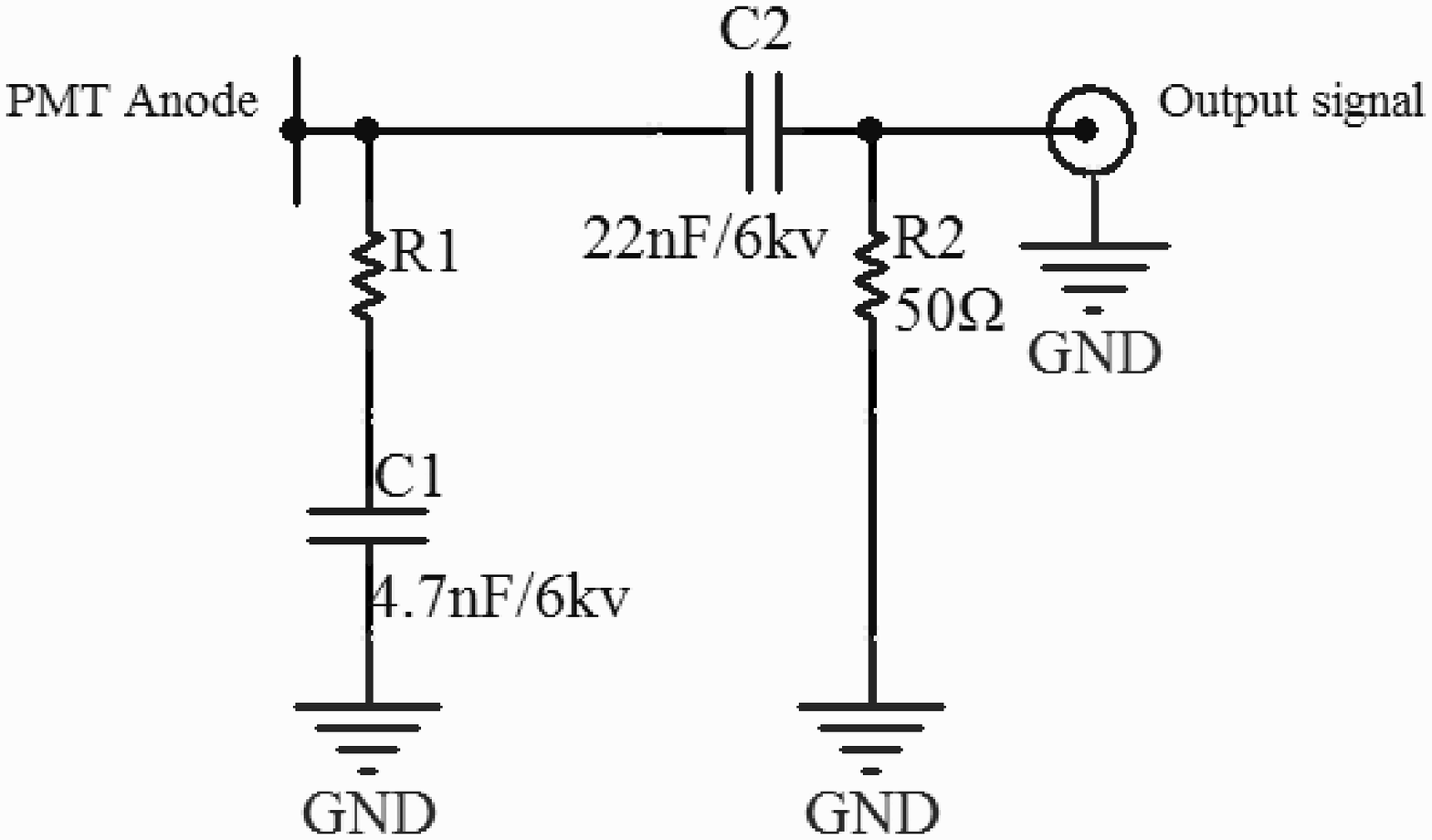}
\figcaption{Simplified circuit of decoupler: differential circuit}
\label{fig4}
\end{center}

 The signal overshoot is a result of the discharging of capacitors with resistors. Here we measured the relationship between the overshoot and R1, where R1 ranged from 1~$\Omega$ to 10~k$\Omega$ following the scheme in Fig.\ref{fig4} ( all other components are shown as in the figure ). The results are shown in Fig.5 and Table~1.
\begin{center}
\label{fig5}
\includegraphics[width=8cm]{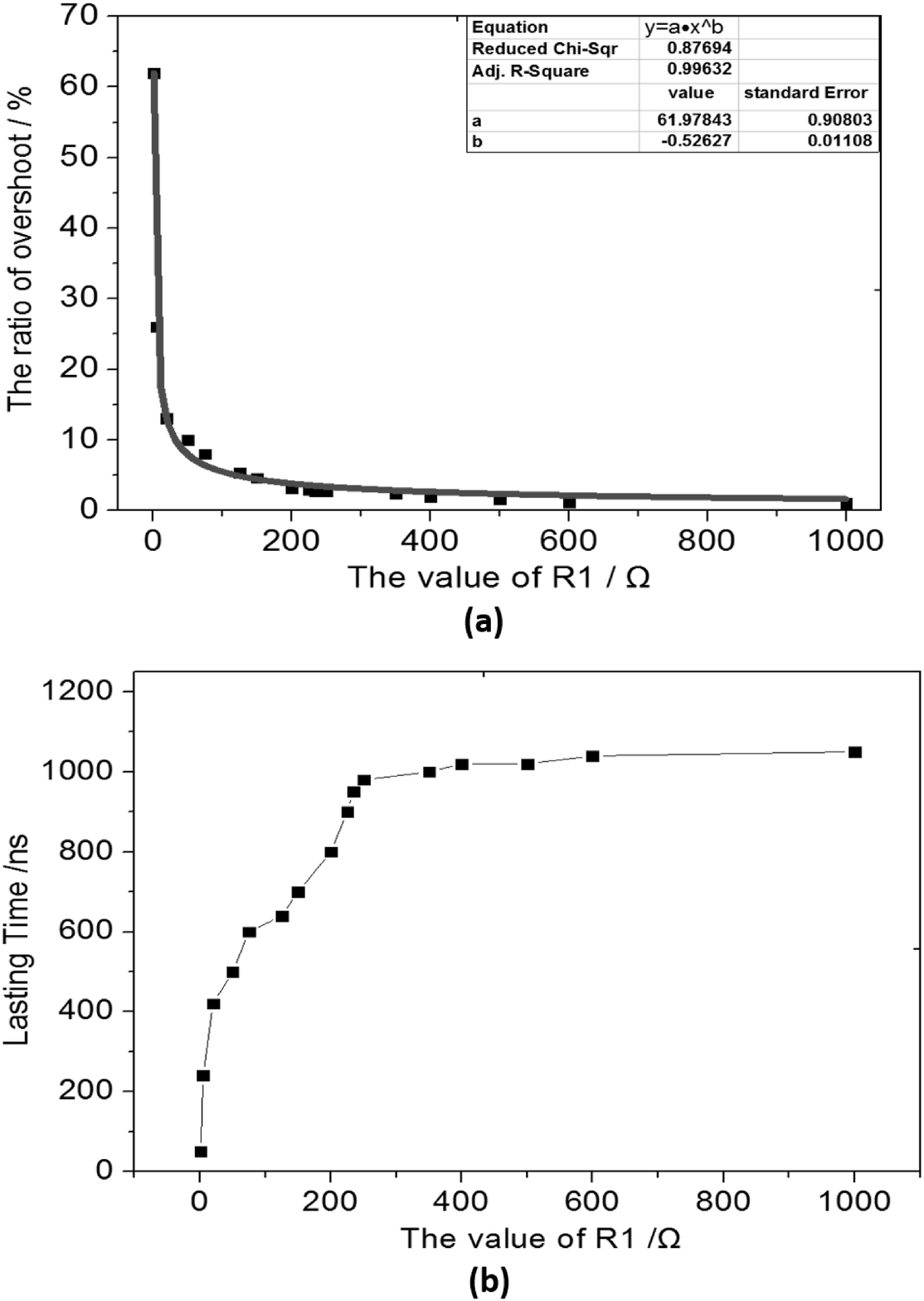}
\figcaption{(a):The ratio of overshoot to signal amplitude as a function of R1; (b): The time of signal baseline recovering to 0.5\% of maximum amplitude of signal as a function of R1.}
\label{fig5}
\end{center}

From Fig.\ref{fig5}(a), we learn that the ratio of overshoot to signal amplitude decreases when R1 increases. According to Fig.\ref{fig5}(b), the time of the signal baseline recovering to 0.5\% of the signal maximum increases when R1 increases where the recovery time is consistent with the time constant of $R1\times C1$. The recovery time shows a knee around R1=250~$\Omega$, above which the overshoot ratio decreases more slowly and the recovery time of the signal baseline almost keeps nearly constant around 1 $\mu$s which is consistent with the time constant of $C2\times R2$. And we tested all 5 types PMTs used in JUNO prototype, and all got consistent results.

From the data ( shown in Table \ref{tab1} ) and calculation, we know that when the value of $R1\times C1<<R2\times C2$, the discharge of $R1\times C1$ contributes more to the overshoot than $R2\times C2$. When $R1\times C1>>R2\times C2$, the discharge of $R2\times C2$ contributes more to the overshoot. We have crosschecked this with different C2 values: when the value is 4.7~nF ( 10~nF,15~nF and 22~nF ), the ratio of overshoot to signal amplitude is around 5\% (2.5\%, 2\% and 1\%), which is consistent with R2 changing under $R2\times C2$.

\begin{center}
\tabcaption{ \label{tab1}  The measurement results of overshoot vs. resistor R1.}
\footnotesize
\begin{tabular*}{80mm}{c@{\extracolsep{\fill}}ccc}
\toprule R1/$\Omega$ & Ratio of overshoot/\%   & Recover time/ns \\
\hline
1\hphantom{000} & \hphantom{0}62 & 50\hphantom{000} \\
5\hphantom{000} & \hphantom{0}26 & 240\hphantom{00} \\
20\hphantom{00} & \hphantom{0}13 & 420\hphantom{00} \\
50\hphantom{00} & \hphantom{0}10 & 500\hphantom{00}\\
75\hphantom{00} & \hphantom{00}8.0 & 600\hphantom{00}\\
125\hphantom{0} & \hphantom{00}5.3 & 640\hphantom{00}\\
150\hphantom{0} & \hphantom{00}4.6 & 750\hphantom{00}\\
200\hphantom{0} & \hphantom{00}3.2 & 800\hphantom{00}\\
225\hphantom{0} & \hphantom{00}2.9 & 900\hphantom{00}\\
234\hphantom{0} & \hphantom{00}2.7 & 950\hphantom{00}\\
250\hphantom{0} & \hphantom{00}2.69 & 980\hphantom{00}\\
350\hphantom{0} & \hphantom{00}2.39 & 1000\hphantom{00}\\
400\hphantom{0} & \hphantom{00}1.95 & 1020\hphantom{00}\\
500\hphantom{0} & \hphantom{00}1.56 & 1020\hphantom{00}\\
600\hphantom{0} & \hphantom{00}1.11 & 1040\hphantom{00}\\
1000\hphantom{0} & \hphantom{00}1.02 & 1100\hphantom{00}\\
10000\hphantom{0} & \hphantom{00}1.0 & 1100\hphantom{00}\\
\bottomrule
\end{tabular*}
\end{center}

\section{Overshoot model}
Following the understanding of overshoot that it is the discharge of the capacitors in the PMT HV divider and decoupler, we can simplify the system further as shown in Fig.\ref{fig6}, where we can model the output of PMT+decoupler as:

\begin{center}
\label{fig6}
\includegraphics[width=6cm]{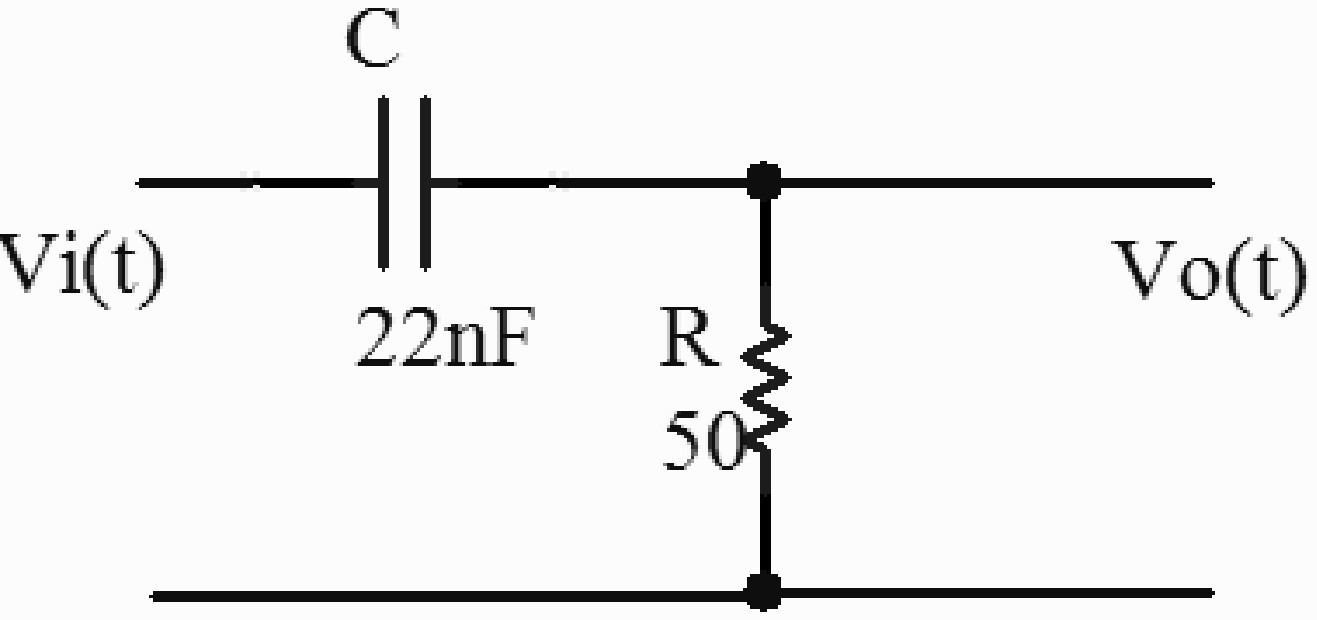}
\figcaption{Further simplified circuit}
\label{fig6}
\end{center}

\begin{equation}
\label{eq1}
V_o(t)=V_i(t) \times h(t)
\end{equation}

Where $V_{i}(t)$ is the PMT output which is the system input, $h(t)$ is the response model of the PMT HV divider + decoupler +oscilloscope and $V_{o}(t)$ is the final waveform viewed by an oscilloscope.

We can use a Fourier transformation to analyze the whole process shown in Fig.\ref{fig7}.
\begin{center}
\label{fig7}
\includegraphics[width=8cm]{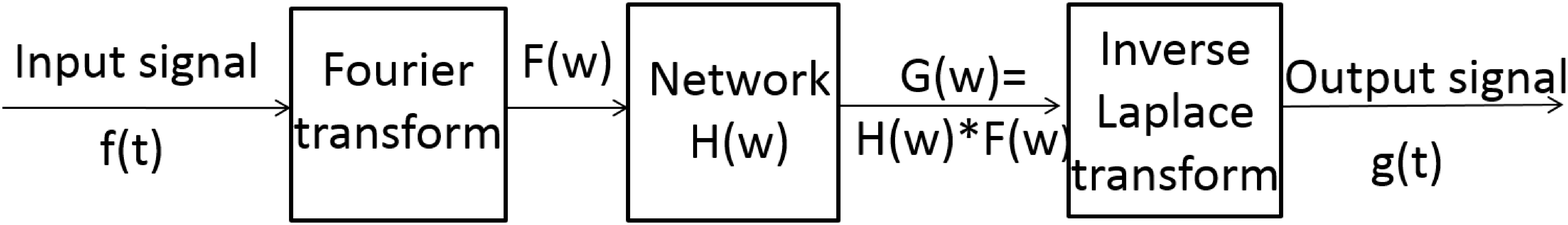}
\figcaption{The signal process with Fourier Transformation analysis. }
\label{fig7}
\end{center}

Adopting the transformation from time to frequency domain, we can get the output signal of the system described as:
\begin{equation}
\label{eq2}
V_o(w)=V_i(w) \times H(w)
\end{equation}
where $H(w)$ is the response of the system in the frequency domain. According to the simplified circuit shown in Fig.\ref{fig6}, it can be expressed as:
\begin{equation}
\label{eq3}
H(w)=jw/(jw+1/{\tau})
\end{equation}
where $\tau$ is the time constant of the differential circuit which is 22~nF$\times$ 50~$\Omega$ $\sim$ 1100~ns in our case.

We can further simplify the input signal $V_{i}(t)$ as an exponential pulse, where $\tau_{i}$ is the time constant of the anode output, which is usually $<$ 12~ns.
\begin{equation}
\label{eq4}
V_i(t)=-Q/C_{i} \times \exp(-t/{\tau}_{i})
\end{equation}

For equivalent calculation, we assume a positive pulse instead of negative pulse. Then we can get the input in the frequency domain,
\begin{equation}
\label{eq5}
V_i(w)=Q/C_{i} \times 1/(jw+1/{\tau}_i))
\end{equation}

Combining equations (\ref{eq1}), (\ref{eq2}), (\ref{eq3}) and (\ref{eq4}), we obtain,
\begin{equation}
\label{eq6}
V_o(w)=V_i(w) \times H(w)=Q/C_{i} \times (jw)/((jw+1/{\tau}) \times (jw+1/{\tau}_i))
\end{equation}

By means of the inverse Fourier transform, the output of system can be obtained:
\begin{equation}
\label{eq7}
V_o(t)=Q/C_i \times (1/({\tau}_i-{\tau})) \times ({\tau}_{i}\exp(-t/{\tau})-{\tau}\exp(-t/{\tau}_{i}))
\end{equation}

Eq.\ref{eq7} is the output of the circuit when the input is an exponential pulse. According to the equation, the output will have an overshoot, and when $\tau>>\tau_{i}$, the overshoot ratio can be expressed as:
\begin{equation}
\label{eq8}
\frac{V_-}{V_M}=\frac{{\tau}_i}{{\tau}}
\end{equation}
where the $V_{-}$ is the amplitude of the overshoot and the $V_{M}$ is the maximum amplitude of the signal. Thus, the value of $V_{-}/V_{M}$ is about 1.1\% for $\tau_{i}$ $\sim$~12~ns and $\tau$ $\sim$1100~ns, or $\sim$8\% for $\tau$ $\sim$140~ns.

For the JUNO prototype, we selected a single ended 50~$\Omega$ matching design with  R1= 10~k$\Omega$ and  C1= 4.7~nF and C2= 22~nF. The final output waveform is shown in Fig.\ref{fig8}. The overshoot of the signal is about 1\% and is consistent with our calculation with the overshoot model.
\begin{center}
\label{fig8}
\includegraphics[width=8cm]{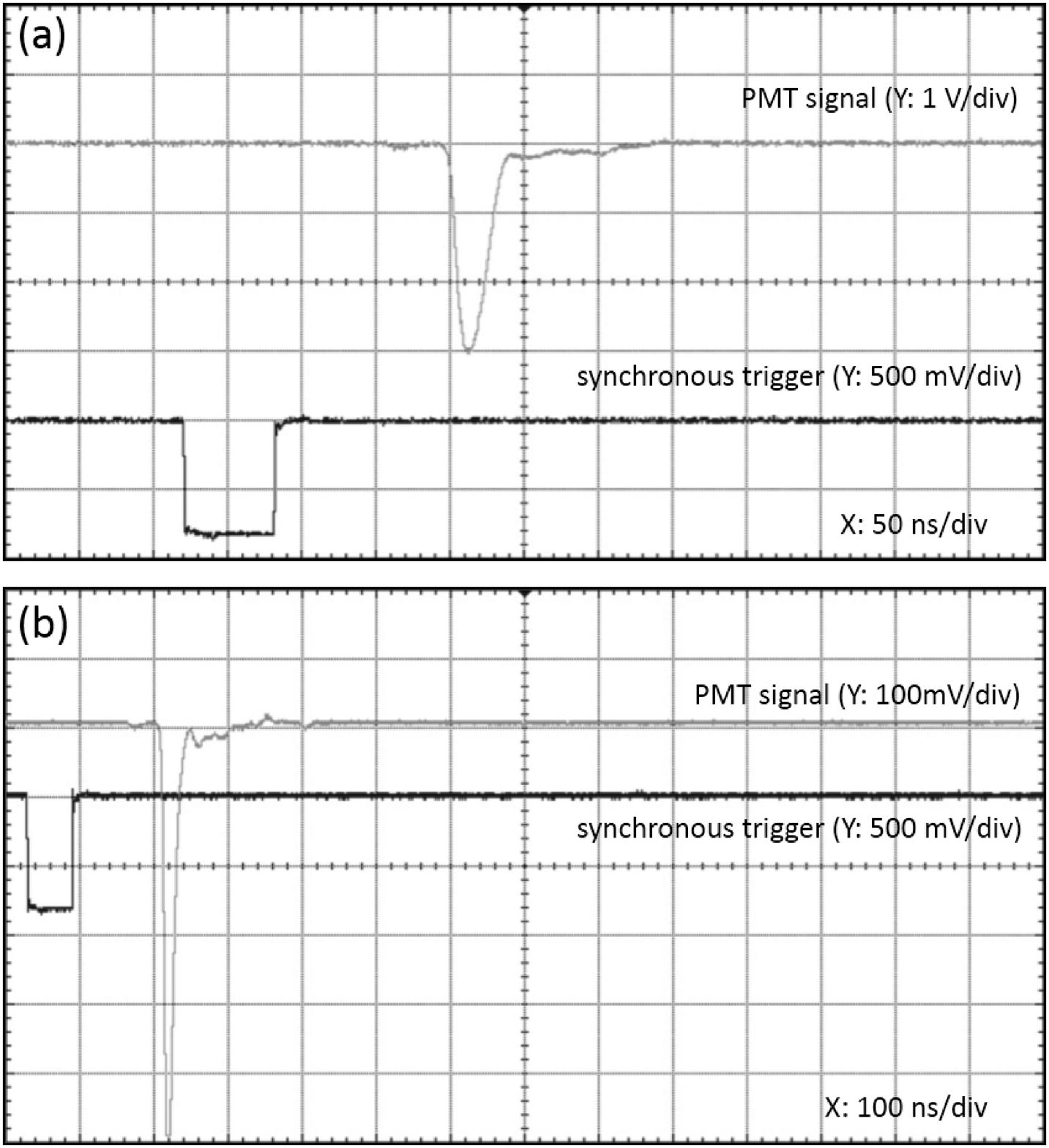}
\figcaption{PMT signal after decoupler: (a)1~V/div and 50~ns/div, (b) the same signal under different scales,100~mV/div and 100~ns/div. The signal amplitude is about 3.0~V. Overshoot is small and the amplitude of overshoot is about 30~mV or 1.0\%, it takes more than 1us to recover.}
\label{fig8}
\end{center}

\section{PMT Performance with updated circuit}

The PMT HV divider plays a crucial role in the PMT performance\cite{lab15}. The timing and linearity of the PMT response are good parameters for checking the PMT HV divider design. We have already reduced the overshoot from $\sim$10\% to $\sim$1\% as discussed in the previous sections. We tested all 5 types PMTs used in JUNO prototype, and all got consistent results. Here we just selected HZC XP1805 PMT as an example to show the typical results.

\subsection{Rise time and fall time}

With the optimized HV divider for a HZC XP1805 PMT, we measured the rise time and fall time with the system shown in Fig.\ref{fig2}.

We get the waveforms of single photoelectron ( SPE ) signals using an LED producing photoelectrons at an average occupancy of 10\%. Then we can record the waveform by oscilloscope with 1~GHz sampling, as shown in Fig.\ref{fig9}.
\begin{center}
\label{fig9}
\includegraphics[width=8cm]{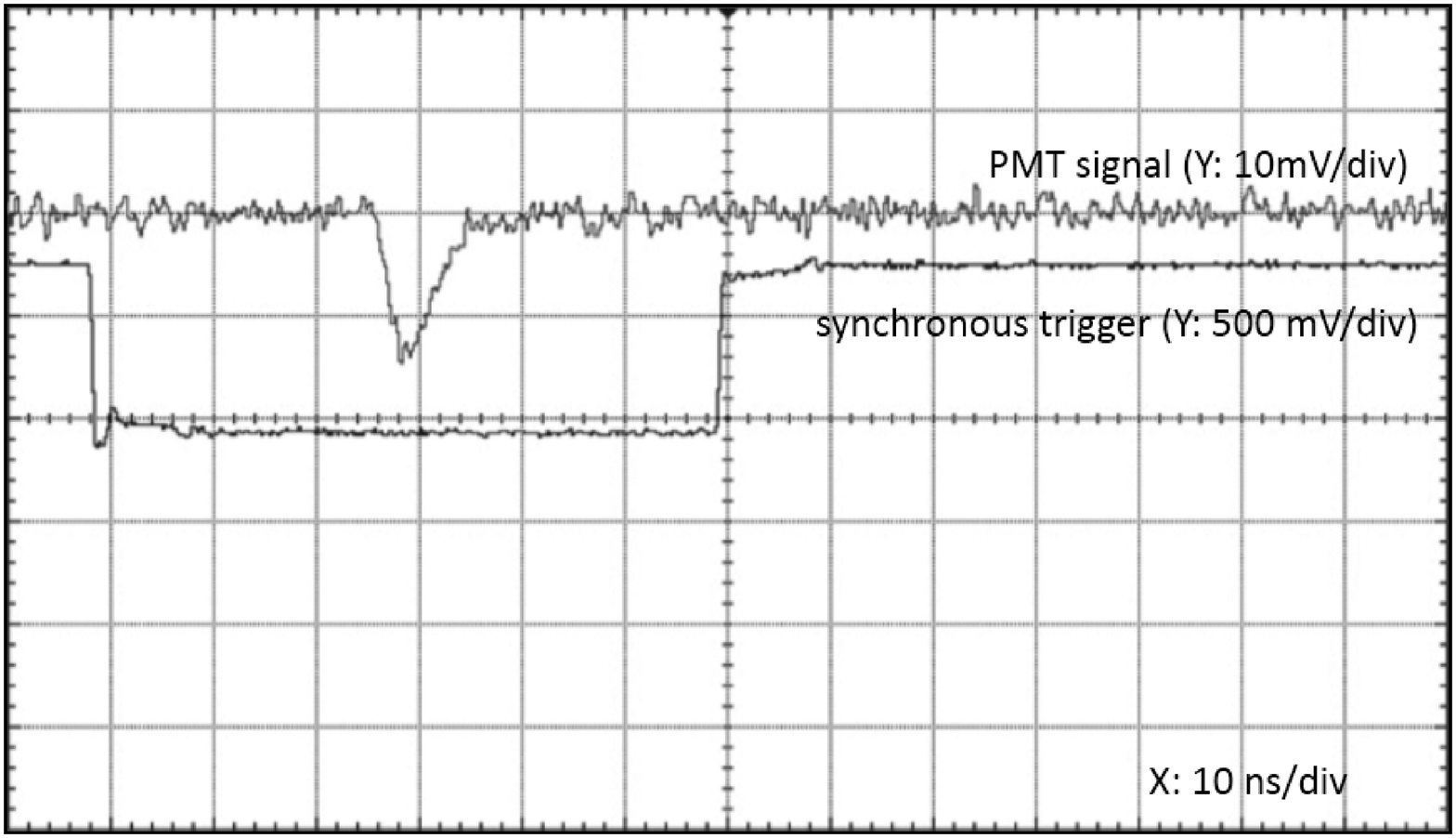}
\figcaption{Captured PMT SPE waveform with scale of 10~mV/div and 10~ns/div, here the gain is 1.2E7, the rise time is 2.2~ns and the fall time is 5.6 ~ns. }
\label{fig9}
\end{center}

From the measured waveform, we know that the PMT with our updated HV divider still has its fast time properties and reaches our expectation.
\subsection{Linearity of the PMT}

The linearity of the PMT can be obviously influenced by the PMT HV divider design. The definition of pulse linearity is the ratio of input and output photoelectrons in pulse operation mode. The measurement scheme is shown in Fig.\ref{fig10}.
\begin{center}
\label{fig10}
\includegraphics[width=8cm]{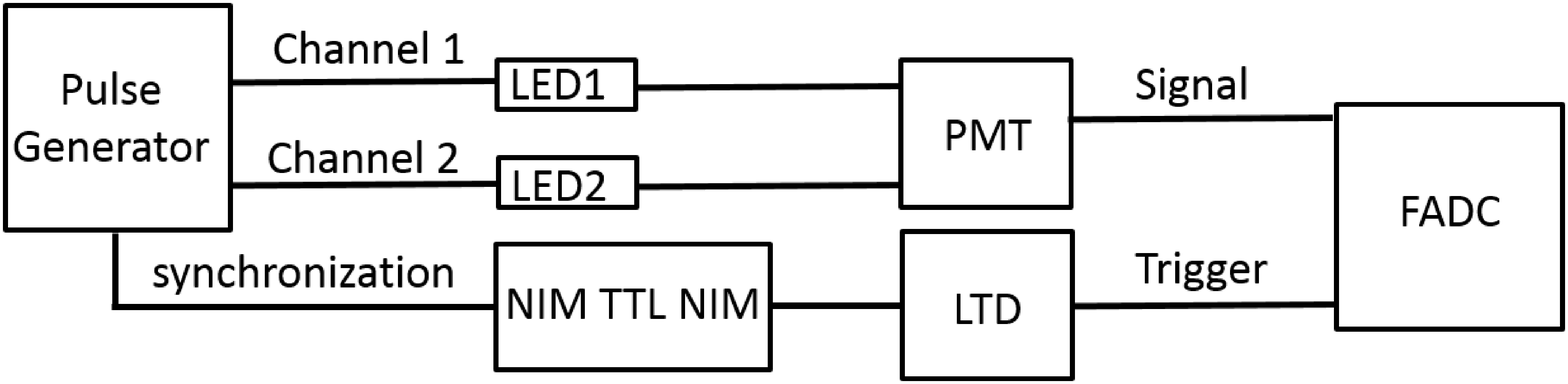}
\figcaption{The schematic of linearity measurement. }
\label{fig10}
\end{center}

A pulse generator can drive two blue LEDs and flash them at the same time. The light intensity of each LED can be tuned separately to cover the whole PMT dynamic range. The PMT can see a sequence of light pulses: a light pulse ( A ) from LED1, a light pulse ( B ) from LED2, and a sum of light pulses ( C ) from the two LEDs flashing simultaneously. At the same time, a FADC system with 1GHz sampling will record the waveforms corresponding to the three different light pulses. If the PMT is ideally linear, we can have,
\begin{equation}\label{3}
C=A+B
\end{equation}

In reality as known, the PMT response has non-linearity effect. Then the deviation of the linearity is defined as,
\begin{equation}
\label{10}
Nonlinearity=\frac{C-(A_{corrected}+B_{corrected})}{A_{corrected}+B_{corrected}}
\end{equation}
Where $A_{corrected}$  ( $B_{corrected}$ ) is the corrected  A ( B )  according to measured  nonlinearity effect in lower intensity range. The measurement results are demonstrated in Fig.\ref{fig11}. The figure shows that the 5\% deviation can reach around 700pe. The measured linearity of the PMT is satisfied our requirements and reached the same level with previous measured R5912\cite{lab16}.

\begin{center}
\label{fig11}
\includegraphics[width=8cm]{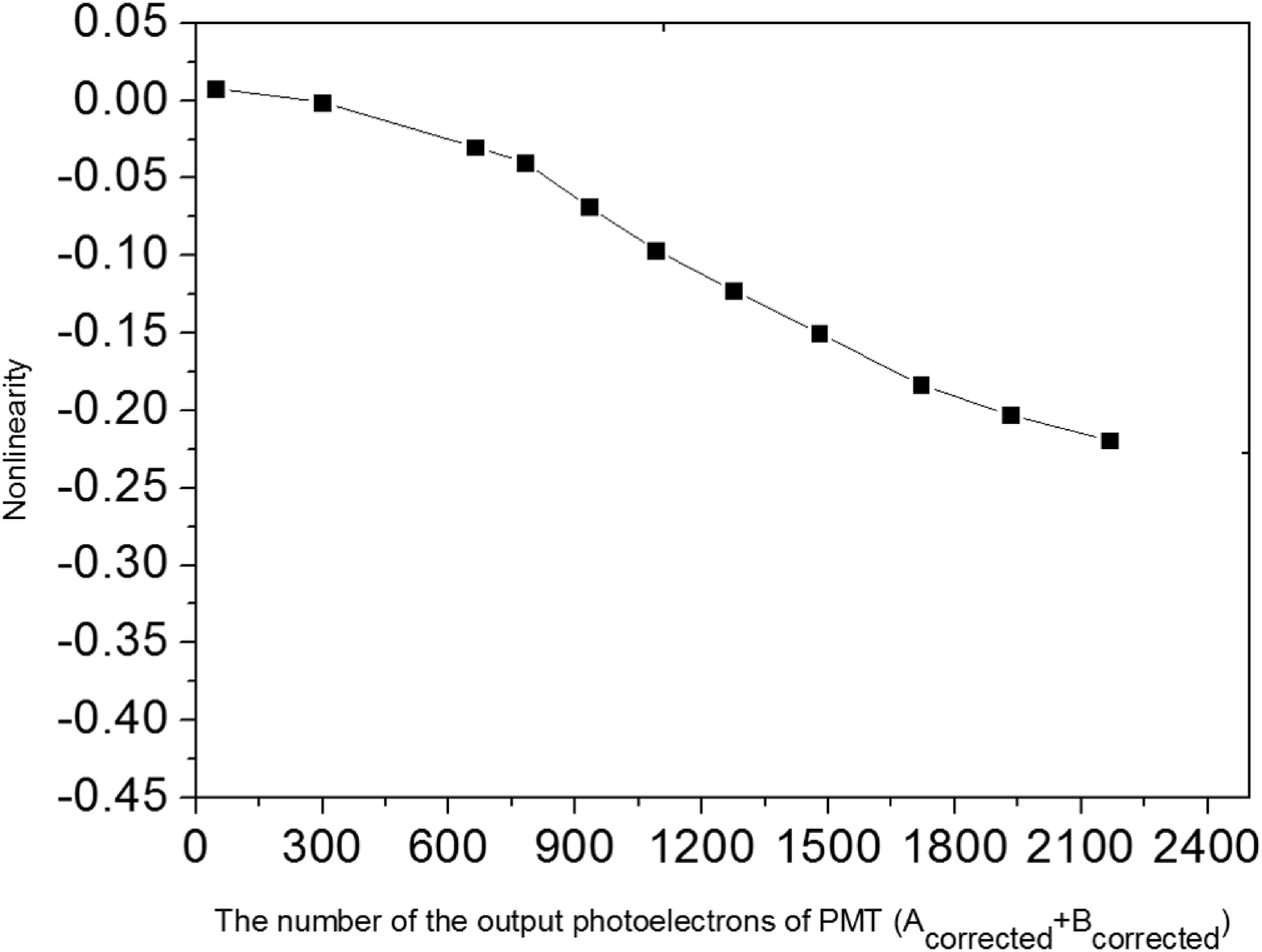}
\figcaption{Nonlinearity as function of the number of the photoelectrons of the anode output pulse. }
\label{fig11}
\end{center}

\section{Conclusion}

In this work, we confirmed a large overshoot of the PMT output at the level of 10\% as observed by the  Daya Bay experiment\cite{lab14}, and made clear that the PMT overshoot originates from the discharge of the decoupling capacitor of the HV-signal decoupler and the capacitor in the HV divider. The design of HV divider and decoupler has been optimized to reduce the overshoot for charge measurements. The value of $R1\times C1$ and $R2\times C2$ are the main contributors to the overshoot. We have also checked the PMT performances with the proposed performance parameters. Further optimization is still needed for the future JUNO HV divider and decoupler to meet other detailed requirements. What's more, a smaller overshoot for a positive voltage PMT is important for charge measurements in a high precision neutrino experiment and will help waveform sampling measurements.
\\

\acknowledgments{ The project was supported by Strategic Priority Research Program A-JUNO, High Energy Physics Experiment and Detector $R\&D$ and National Natural Science Foundation of China.}

\end{multicols}

\vspace{-1mm}
\centerline{\rule{80mm}{0.1pt}}
\vspace{2mm}

\begin{multicols}{2}

\end{multicols}

\clearpage
\end{CJK*}
\end{document}